\documentclass[twoside,a4paper,11pt]{sea9}
\usepackage{graphicx}
\usepackage{hyperref}
\usepackage{movie15}
\usepackage[utf8]{inputenc}

\begin{document}
\pagenumbering{arabic}
\pagestyle{myheadings}
\thispagestyle{empty}
{\flushleft\includegraphics[width=\textwidth,bb=58 650 590 680]{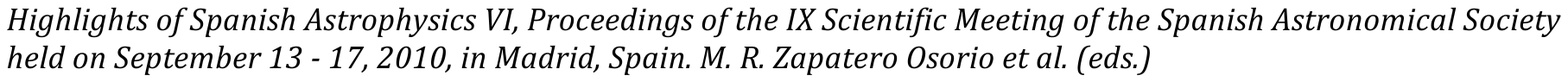}}
\vspace*{0.2cm}
\begin{flushleft}
{\bf {\LARGE
%
Spin period evolution of the X-ray pulsar GX 1+4
%
}\\
\vspace*{1cm}
%
A. González-Galán$^{1}$,
E. Kuulkers$^{2}$, 
P. Kretschmar$^{2}$,
S. Larsson$^{3}$,
K. Postnov$^{4}$,
A. Kochetkova$^{4}$,
and 
M. H. Finger$^{5,6}$
%
}\\
\vspace*{0.5cm}
%
$^{1}$
Departamento de F\'isica, Ingenier\'ia de Sistemas y Teor\'ia de la Se\~nal, University of Alicante, P.O.~Box 99 E-03080 Alicante, Spain\\
$^{2}$
European Space Agency, European Space Astronomy Centre, P.O.~Box 78, 28691, Villanueva de la Cañada, Madrid, Spain\\
$^{3}$
Department of Astronomy, Stockholm University, SE-106 91 Stockholm, Sweden\\
$^{4}$
Sternberg Astronomical Institute, 119992, Moscow, Russia\\
$^{5}$
National Space Science and Technology Center, 320 Sparkman Drive, Huntsville, AL 35805, USA\\
$^{6}$
Universities Space Research Association, 6767 Old Madison Pike, Suite 450, Huntsville, AL 35806, USA\\
%
\end{flushleft}
%
\markboth{
Spin period evolution of GX 1+4
}{ 
%
González-Galán, A. et al.
%
}
\thispagestyle{empty}
\vspace*{0.4cm}
\begin{minipage}[l]{0.09\textwidth}
\ 
\end{minipage}
\begin{minipage}[r]{0.9\textwidth}
\vspace{1cm}
\section*{Abstract}{\small
%
We report on the long-term evolution of the spin period of the symbiotic X-ray pulsar GX~1+4 and a possible interpretation within a model of quasi-spherical accretion. New period measurements from BeppoSAX, INTEGRAL and  Fermi observations have been combined with previously published data from four decades of observations. During the 1970's GX~1+4 was spinning  up with the fastest rate among the known X-ray pulsars at the time. In the mid 1980's it underwent a change during a period of low X-ray flux and started to spin down with a rate similar in magnitude to the previous spin up rate. The spin period has changed from $\sim$110~s to $\sim$160~s within the last three decades. Our results demonstrate that the overall spin down trend continues and is stronger than ever. We compare the observations with predictions from a model assuming quasi-spherical accretion from the slow wind of the M giant companion. 
%
\normalsize}
\end{minipage}
%
%
%
\section{Introduction \label{intro}}

Accreting X-ray pulsars are highly magnetized neutron stars in a binary system, accreting matter from their companion star. The mass transfer can take place via Roche-Lobe overflow, strong stellar winds for giant stars or the Be emission mechanism. These accreting pulsars radiate predominantly in the X-ray band, and the radiation is modulated by the stellar rotation of the pulsar. For a review on this subject, see e.g., Nagase~\cite{nagase89}.

GX~1+4 is an accreting X-ray pulsar which was discovered in 1970 by a balloon X-ray observation at energies above 15 keV showing pulsations with a period of about two minutes (Lewin~et~al.~\cite{lewin71}). It was one of the brightest X-ray sources in the Galactic center region. The composite emission spectrum of GX~1+4 indicated that the object was almost certainly a binary system, consisting of a symbiotic red giant and a much hotter source (Davidsen~et~al.~\cite{davidsen77}). Chakrabarty~\&~Roche~\cite{roche97} confirmed the optical  companion to be the M giant V2116 Oph.

More recently, observations have shown a well-determined, 1161 day period, single-line spectroscopic binary orbit (Hinkle~et~al.~\cite{hinkle2006}). The inclination of the orbit is, however, unknown. Therefore, on the basis of similar characteristics seen in other X-ray binary systems the mass of the neutron star is inferred to be $\sim1.35$ $M_\odot$. The mass function of the system, combined with the abundances of the M giant star and the assumed mass for the neutron star, indicate a mass of $\sim1.2$ $M_\odot$ for the M giant star. This implies that the M giant star is a first ascent giant which does not fill its Roche Lobe. In consequence, the neutron star is capturing the slow stellar wind of its companion, which makes V2116 Oph quite different from other low-mass X-ray binaries  (Hinkle~et~al.~\cite{hinkle2006}). Therein, GX~1+4 is the first low mass X-ray binary with an M giant optical companion, and the first symbiotic X-ray pulsar ever observed.

\section{Results}

\begin{figure}
\center
\vspace{-1.5cm}
\includegraphics[width=0.9\textwidth]{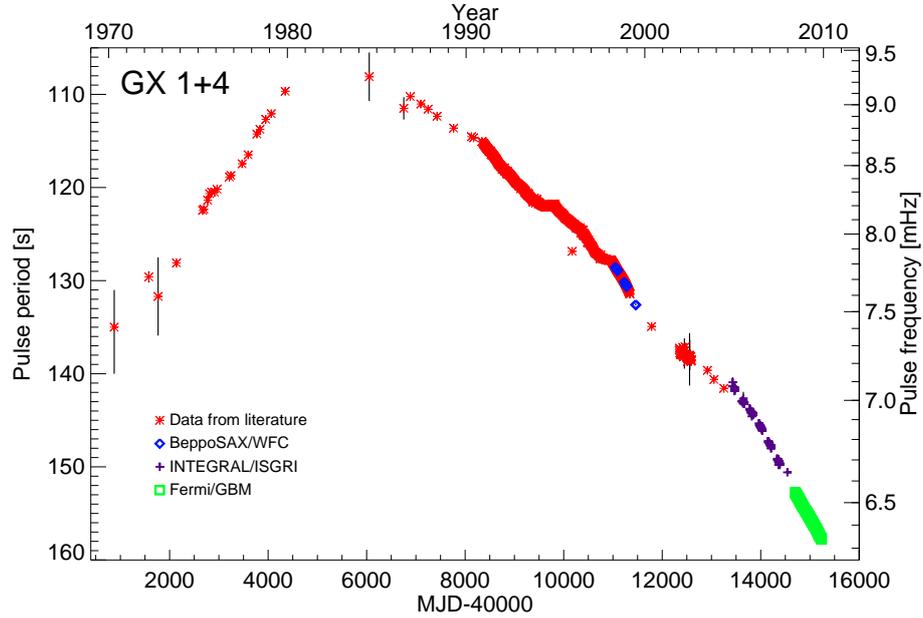} 
\vspace{-1.5cm}
\caption{\label{GXhistdata}Long-term pulse period of GX~1+4 (González-Galán et al. \cite{gonzalez2010} and references therein). Note that the period increases from top to bottom.}
\end{figure}

\begin{figure}
\center
\vspace{-0.5cm}
\includegraphics[width=0.45\textwidth,angle=180]{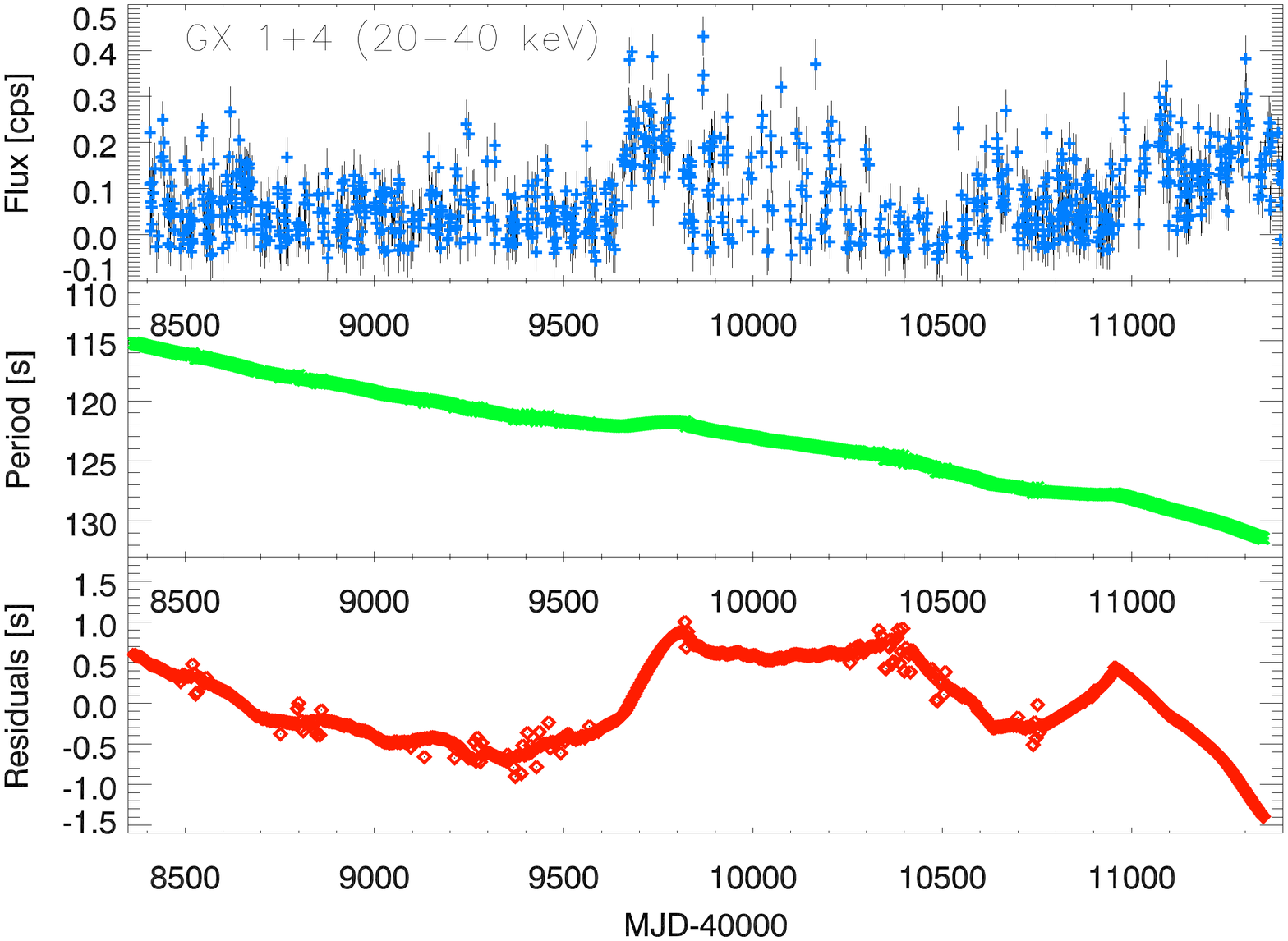}~
\includegraphics[width=0.45\textwidth,angle=180]{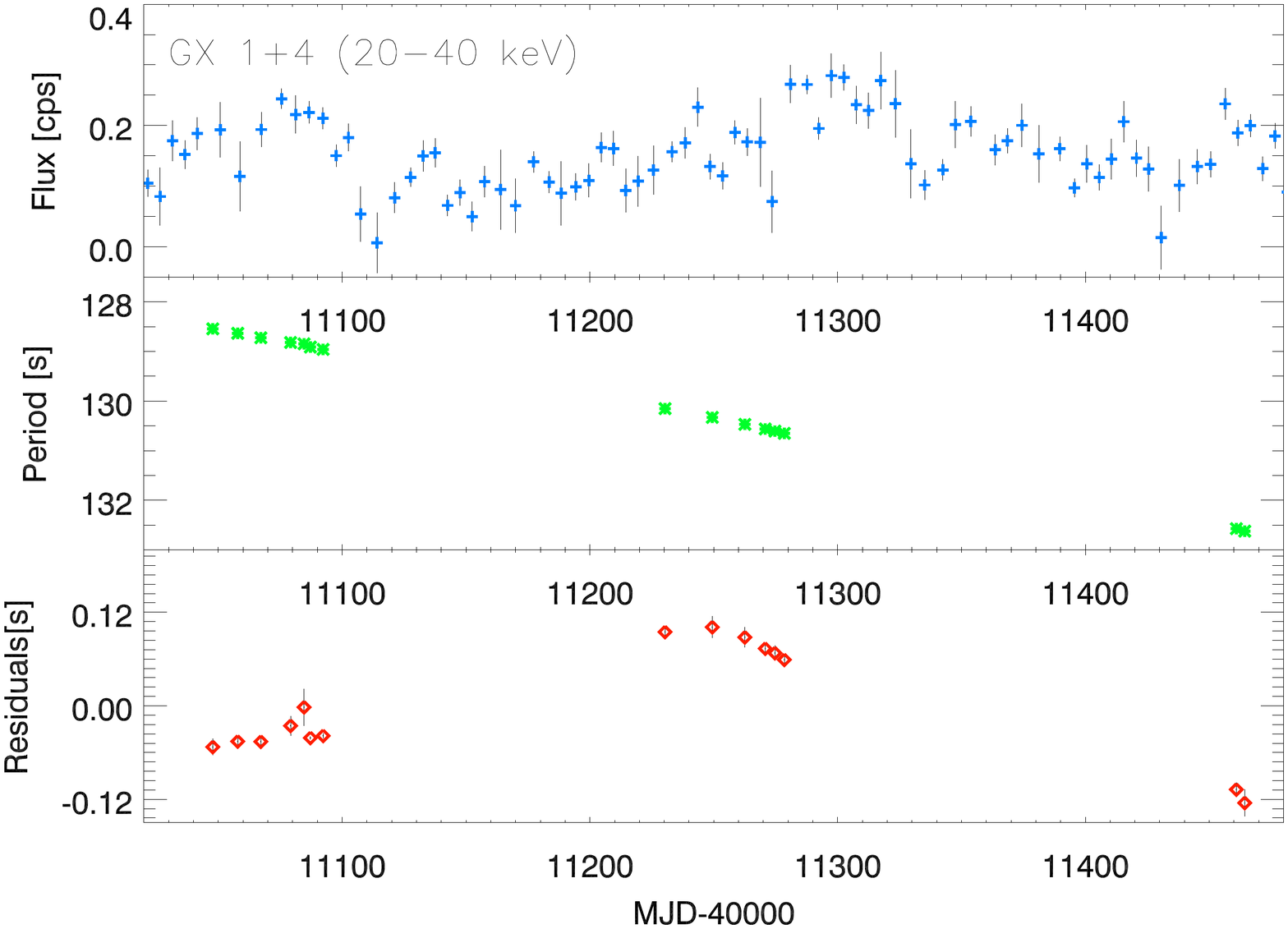} \\
\vspace{-0.5cm}
\includegraphics[width=0.45\textwidth,angle=180]{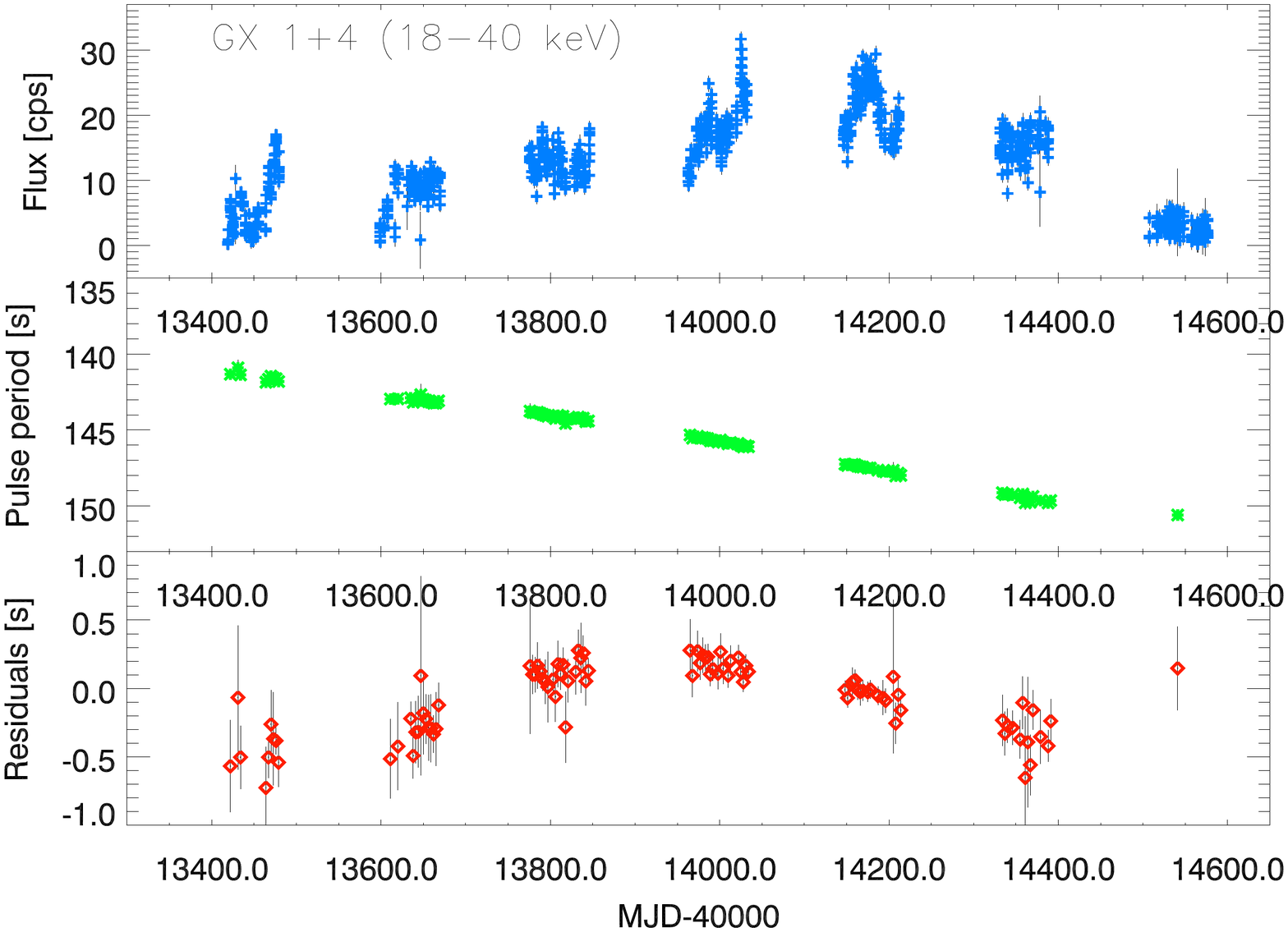} ~
\includegraphics[width=0.45\textwidth,angle=180]{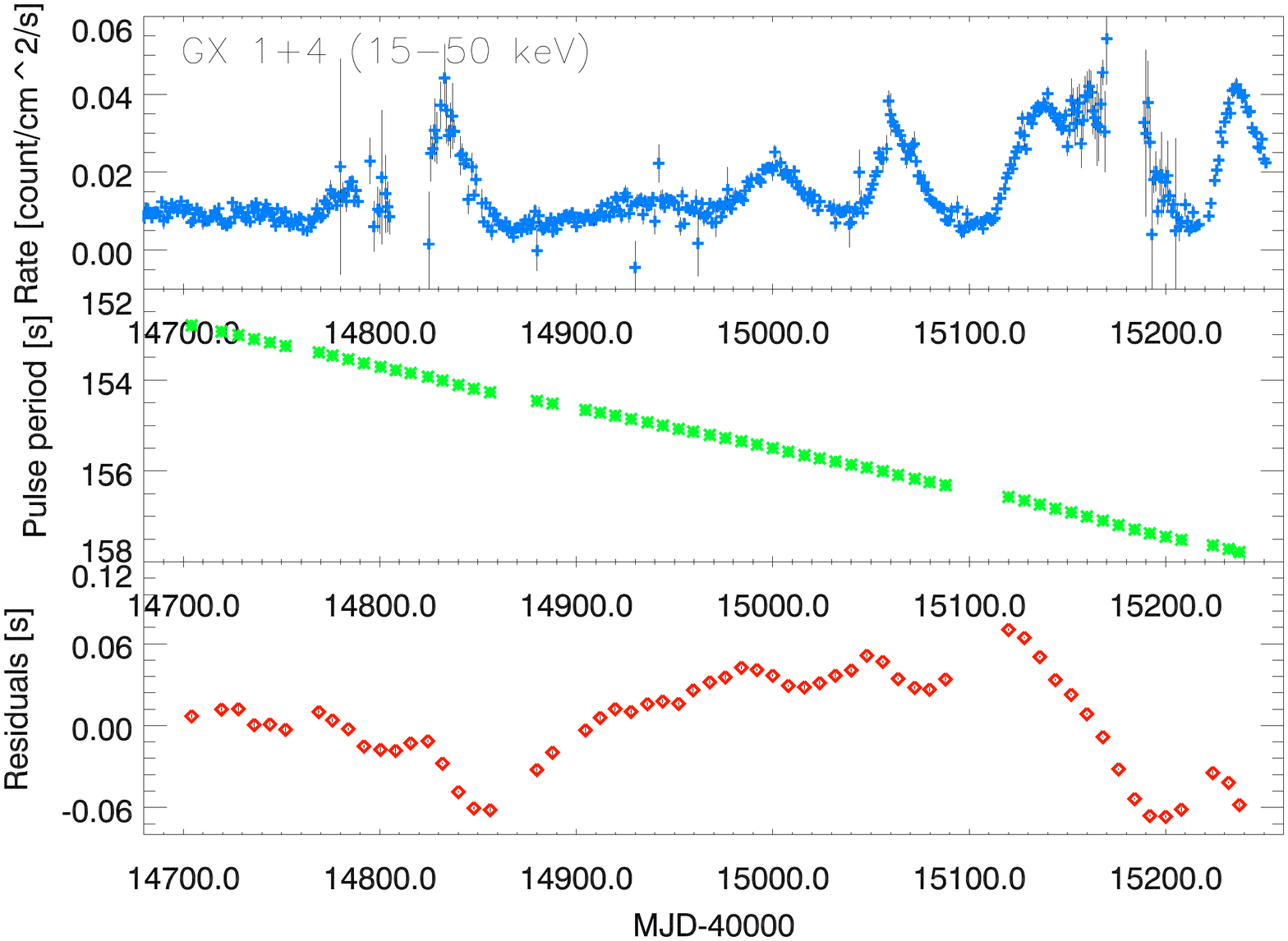} 
\vspace{-0.1cm}
\caption{\label{fluxesperiods} GX~1+4 data. \scriptsize{{\bf Top left plot:} {\it Top:} CGRO/BATSE daily lightcurve in the energy range 20-40 keV. {\it Middle:} pulse periods derived from CGRO/BATSE data. {\it Bottom:} Residuals of the periods from a linear fit. {\bf Top right plot:} {\it Top:} CGRO/BATSE daily lightcurve in the energy range 20-40 keV. {\it Middle:} pulse periods derived from BeppoSAX/WFC data. {\it Bottom:} Residuals of the periods from a linear fit. {\bf Bottom left panel:} INTEGRAL/ISGRI data. {\it Top:} Average flux per pointing in the energy range 18 to 40 keV. {\it Middle:} 20-40 keV pulse period. {\it Bottom:} Residuals from the pulse periods using a linear fit. {\bf Bottom right plot:} {\it Top:} Swift/BAT daily averaged light curve in the energy range 15--50 keV. {\it Middle:} Pulse periods derived from Fermi/GBM data. {\it Bottom:} Residuals from the pulse periods using a linear fit. Note that in all these plots the period increases from top to bottom.}
}
\vspace{-0.6cm}
\end{figure}

The main result obtained during this work is the plot shown in Fig.~\ref{GXhistdata} which currently is the most updated and complete series of pulse periods of GX~1+4. The available measurements of the pulse period of the neutron star of GX~1+4 span a time period of about 40 years giving us a unique insight in the pulse period evolution of this symbiotic X-ray pulsar. To obtain this result we have combined all measurements that we are aware of from the literature with the pulse periods determined from BeppoSAX/WFC, INTEGRAL/ISGRI and Fermi/GBM data.

The system has a peculiar spin history. During the 1970's GX~1+4 was spinning up with the fastest rate ($\dot{P}\sim7.23 \times 10^{-8}$~s/s) among the known X-ray pulsars at the time (see e.g., Nagase~\cite{nagase89}). During that period the source had an X-ray luminosity between $9 \times 10^{38}$ and $2.25 \times 10^{38}$~erg~s$^{-1}$ (White~et~al.~\cite{white83}) (the luminosity value have been corrected to the distance range (3-6~kpc) (Chakrabarty~\&~Roche~\cite{roche97}) ). However, EXOSAT observations in 1983 revealed an extended low state with an X-ray luminosity of $< 4 \times 10^{36}$~erg~s$^{-1}$ (Hall~\&~Davelar~\cite{hall83}). Around the same time GX~1+4 started to spin down with a rate similar in magnitude to the previous spin-up rate (see Fig.~\ref{GXhistdata}) and continues spinning down up to date. Indeed, the spin down rate measured by Fermi is the strongest spin down rate observed to date ($\dot{P}\sim1.08 \times 10^{-7}$~s/s), and the pulse period of the pulsar has increased by about 50\% during the last $\sim30$ years to the largest value ever known for this source ($\sim160$~s) (González-Galán~et~al.~\cite{gonzalez2010}). 

Apart from the long-term plot, looking at the behavior along shorter time scales (Fig.~\ref{fluxesperiods}), it is possible to observe that even when the X-ray flux has variations, pulse period evolution seems to be almost linear.

The unusual long-term spin behavior of GX~1+4 has attracted considerable interest for many years. Studying the pulse period evolution in an accreting X-ray pulsar and relating it to, e.g., the luminosity changes, allows to test models of accretion and to gain insights on the physical processes taking place in this kind of systems. In González-Galán~et~al.~\cite{gonzalez2010} we extend the investigation of the spin period history of GX~1+4 with new observations obtained by BeppoSAX, INTEGRAL and Fermi, we correlate X-ray flux variations with pulse period variations of the system using the same observations as well as those obtained with Swift and BATSE, and provide a likely explanation for the spin history seen.

\section{Accretion models}
\label{theoretical}

Standard accretion theory by Ghosh~\&~Lamb~\cite{ghoshlamb79} assumes the formation of a prograde disk around the neutron star. This prograde disk accelerates the neutron star through the angular momentum of the accreted matter. Therefore, spin down of the pulsar is only possible in the quasi-equilibrium state, and there is a positive correlation between accretion rate ($\dot{M}$) and the acceleration of the neutron star. Assuming $\dot{M} \propto F_{X}$, the positive correlation $\dot{\nu} \propto F_{X} $ should be observed in the data even during spin down.

Transient disks with an alternating sense of rotation are known to form in numerical simulations in binary systems fed from stellar wind (Fryxell~\&~Taam~\cite{frixell88}), therefore it is also possible the formation of a transient retrograde disk in GX~1+4 as it is a wind-fed source. Indeed, a retrograde accretion disk was proposed for the first time by Makishima~et~al.~\cite{makishima88} to explain the spin down observed in the neutron star of GX~1+4. This model assumes a retrograde disk around the neutron star decelerating the pulsar through the angular momentum of accreted matter, therein, a negative correlation $-\dot{\nu} \propto F_{X}$ should be observed in the data.

There is a third model proposed for the first time to explain the behavior of GX~1+4 in González-Galán~et~al.~\cite{gonzalez2010}, the quasi-spherical accretion model. This model predicts a quasi-static atmosphere around the neutron star (instead of a disk) which carries away angular momentum from the neutron star magnetosphere but allows accretion (in contrast to the so-called subsonic propeller regime (Davies \& Pringle\cite{davies80})); and a negative correlation ($-\dot{\nu} \propto F_{X}$) during spin down (see Shakura~et~al.~\cite{shakura2011}; Postnov~et~al.~\cite{postnov2010}). The formation of this kind of "atmosphere" is only possible for wind-fed sources like GX~1+4.

\section{Discussion}

\begin{figure}
\center
\vspace{-2.7cm}
\includegraphics[width=0.7\textwidth]{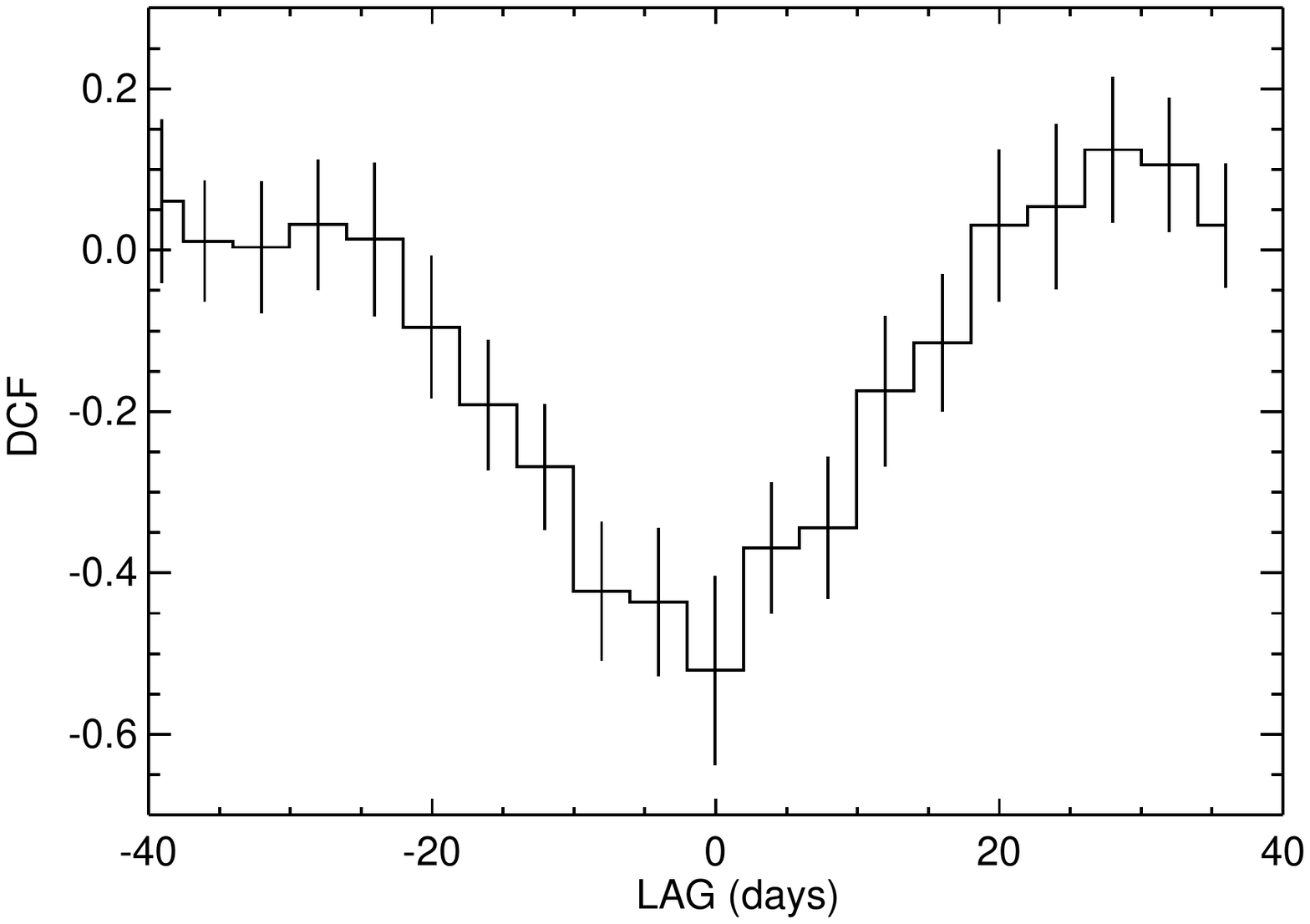}
\vspace{-6.7cm}
\caption{\label{crosscorr}The discrete correlation function between $\dot \nu$ obtained from Fermi/GBM GX~1+4 data and the Swift/BAT 15 - 50 keV X-ray flux shown in bottom right panel of Fig.~\ref{fluxesperiods}. The minimum near zero lag implies a negative correlation between the pulse frequency derivative and the X-ray flux (González-Galán~et~al. \cite{gonzalez2010}).}
\end{figure}

The correlations between X-ray flux and pulse period evolution predicted by the models proposed in Section~\ref{theoretical} can be tested by comparison with a regular sequence of X-ray fluxes and pulse periods which can be found in BATSE and Fermi data (see Fig.~\ref{fluxesperiods}) for GX~1+4.

A cross-correlation of $\dot\nu$ obtained from Fermi/GBM data and the contemporaneous X-ray flux measured by Swift/BAT finds a minimum near zero lag (see Fig.~\ref{crosscorr}) implying a negative correlation between this two parameters for GX~1+4. Furthermore, this negative correlation has already been found previously in BATSE data (i.e. Chakrabarty~et~al.~\cite{chakrabarty97}). This negative correlation discards the standard accretion disk model as a possible interpretation of GX~1+4 behavior.

Due to several reasons, including this negative correlation, different authors have tried to explain GX~1+4 behavior assuming the formation of a retrograde disk around the neutron star by matter captured from stellar wind of the secondary M-giant star in GX~1+4 (e.g. Makishima~et~al.~\cite{makishima88}, Dotani~et~al.~\cite{dotani89}, Chakrabarty~et~al.~\cite{chakrabarty97}, Nelson~et~al.~\cite{nelson97}, Ferrigno~et~al.~\cite{ferrigno2007}, etc.). The weak points of the retrograde disk interpretation, however, include the puzzling long-term ($\sim30$ years) stability of such a retrograde disk.

This puzzling long-term stability of the retrograde disk interpretation has led us to a new model, the quasi-spherical accretion model, which is able to explain the negative correlation found in the data without the formation a retrograde accretion disk (González-Galán~et~al.~\cite{gonzalez2010}).

\section{Conclusions}

New measurements by INTEGRAL and Fermi confirm the overall spin down of the neutron star in GX 1+4 observed over the last decades. This spin down is stronger than ever observed before with an increasement of the spin period of $\sim50\%$ during the last $\sim30$ years (González-Galán~et~al.~\cite{gonzalez2010}). The analysis of Fermi data reveals a negative correlation between spin frequency derivative and the X-ray flux simultaneously measured by Swift/BAT monitor (González-Galán~et~al.~\cite{gonzalez2010}), which confirms the correlation of instantaneous spin-down torque with X-ray flux discovered by BATSE (Chakrabarty~et~al.~\cite{chakrabarty97}) discarding clearly the standard accretion disk model for this system. We have shown that in GX 1+4 not only the retrograde disk accretion is a possible explanation but the quasi-spherical accretion onto the neutron star from the stellar wind of the M giant companion is likely to take place (González-Galán~et~al.~\cite{gonzalez2010}). 

%
%
\small  
%
\section*{Acknowledgments}   
%
Partly based on observations with INTEGRAL, an ESA project with instruments and science data centre funded by ESA member states (especially the PI countries: Denmark, France, Germany, Italy, Switzerland, Spain), Czech Republic and Poland, and with the participation of Russia and the USA. We acknowledge support from the Faculty of the European Space Astronomy Centre (ESAC), and from NASA grant NNX08AW06G. The work of KP and AK was partially supported by RFBR grant 10-02-00599. AG is funded by the Spanish MICINN under grant AYA2008-06166-C03-03 and belongs to the Consolider-Ingenio 2010-GTC Program (grant CSD2006-70) of the Spanish MICINN.The Swift/BAT transient monitor results are provided by the Swift/BAT team. We thank Deepto Chakrabarty for providing most of the historical pulse measurements and Jean in 't Zand for providing the BeppoSAX/WFC light curves.
%

%
\end{document}